\documentclass{article}
\usepackage{kr}

\usepackage{times}
\usepackage{soul}
\usepackage{url}
\usepackage[hidelinks]{hyperref}
\usepackage[utf8]{inputenc}
\usepackage[small]{caption}
\usepackage{graphicx}
\usepackage{amsmath}
\usepackage{amsthm}
\usepackage{booktabs}
\usepackage[linesnumbered,ruled,vlined]{algorithm2e}

\newtheorem{example}{Example}

\title{A Distributed Framework for Compiling and Reasoning with d-DNNF}

\author{%
Zhenghang Xu$^{1, 2}$\and
Minghao Yin$^{1, 2}$ \and
Jianan Wang$^{1}$ \and
Jean-Marie Lagniez$^3$ \\
\affiliations
$^1$School of Information Science and Technology, Northeast Normal University, Changchun, China \\
$^2$Key Laboratory of Applied Statistics of MOE, Northeast Normal University, Changchun, China \\
$^3$Univ. Artois, CNRS, CRIL, F-62300 Lens, France\\
\emails
\{xuzh121, ymh, wangjn\}@nenu.edu.cn,
lagniez@cril.fr
}

\usepackage{xspace}

\def\discount{{\tt DisCount}}

\def\d4{{\tt D4}}

\def\dcountAntom{{\tt dCountAntom}}
\def\dmc{{\tt DMC}}

\def\dkc{{\tt dkc}}
\def\dreasoner{{\tt dreasoner}}

\usepackage{mathtools}
\DeclarePairedDelimiterX{\norm}[1]{\lVert}{\rVert}{#1}

\newcommand{\dfour}{{\tt d4}\xspace}

\newcommand{\ctwod}{{\tt C2D}\xspace}
\newcommand{\dSharp}{{\tt Dsharp}\xspace}
\newcommand{\sharpSAT}{{\tt SharpSAT-TD}\xspace}

\newcommand{\cnf}{CNF\xspace}

\newcommand{\dDNNF}{d-DNNF\xspace}
\newcommand{\decDNNF}{decision-DNNF\xspace}

\begin{document}

\maketitle

\begin{abstract}
Knowledge Compilation (KC) is a powerful paradigm that enables efficient reasoning by transforming propositional formulas into tractable target languages, such as Deterministic, Decomposable Negation Normal Form (d-DNNF). 
However, as real-world problem instances grow in complexity, the offline compilation phase becomes a significant computational bottleneck, often exceeding the memory and temporal limits of single-node systems. 
While distributed computing has been successfully applied to model counting ($\#\mathsf{SAT}$), extending these techniques to knowledge compilation remains a challenge due to the difficulty of sharing partial circuit fragments across distributed nodes. 

In this paper, we propose \dkc{}, the first distributed knowledge compiler designed for large-scale Decision-DNNF generation.
Leveraging a Cube-and-Conquer strategy, \dkc{} effectively partitions the search space into independent subproblems, mitigating the communication overhead typically associated with work-stealing architectures in circuit-based tasks. 
Recognizing that the utility of compilation lies in subsequent querying, we further introduce \dreasoner{}, a distributed reasoning engine. 
\dreasoner{} is capable of performing core inference tasks (including model counting, direct access, and uniform sampling) across a distributed d-DNNF structure, even under variable conditioning. 
Our experimental evaluation on benchmarks demonstrates that our distributed architecture scales effectively, enabling the compilation and querying of complex formulas that remain beyond the reach of state-of-the-art sequential compilers.
\end{abstract}

\section{Introduction~\label{intro}}
Propositional logic serves as a cornerstone of Artificial Intelligence, providing a rigorous framework for knowledge representation and reasoning~\cite{DBLP:books/daglib/0023546}. 
It underpins critical subfields, including Explainable AI (XAI)~\cite{DBLP:conf/lics/Darwiche23}, automated reasoning~\cite{DBLP:series/faia/336}, and decision support systems~\cite{DBLP:conf/cade/Plaisted15}. 
However, the expressiveness of propositional logic comes at a steep computational cost: core inference tasks, such as satisfiability (SAT), are $\mathsf{NP}$-complete~\cite{DBLP:conf/stoc/Cook71}, while counting tasks like $\#\mathsf{SAT}$ reside in the complexity class $\#\mathsf{P}$. 
This inherent complexity creates a bottleneck for real-time applications, particularly in high-stakes environments where low-latency reasoning is non-negotiable.

Industrial applications exacerbate this challenge, demanding scalability for massive problem instances. 
For example, AWS’s \emph{Zelkova} engine~\cite{DBLP:conf/cav/Rungta22} processes billions of verification queries daily, highlighting the urgent need for optimized reasoning pipelines. 
Beyond simple satisfiability, practical tasks such as \emph{model counting}, \emph{direct access}, \emph{uniform sampling}, and \emph{model enumeration} are essential for extracting meaningful insights from complex constraints.

To mitigate online computational costs, \emph{Knowledge Compilation} (KC) has emerged as a standard paradigm~\cite{DBLP:journals/aicom/CadoliD97,DBLP:journals/amai/DarwicheM24}. 
KC shifts the computational burden to an offline phase, transforming propositional formulas into target languages that support polynomial-time queries. 
Prominent among these is the \emph{Deterministic, Decomposable Negation Normal Form} (d-DNNF) language, which renders tasks like Weighted Model Counting (WMC) tractable. 
The practical utility of this approach is evident in systems like Renault’s configuration tool~\cite{DBLP:conf/ictai/AstesanaCF10}, where precomputed compilations enable rapid, interactive product customization.

Despite its benefits, compiling large-scale formulas into d-DNNF remains a significant hurdle; the process is both time and memory intensive, often exceeding the resources of a single machine for real-world instances. 
To handle this growing complexity, parallel and distributed computing offers a promising path forward. 
Given the structural proximity between $\#\mathsf{SAT}$ solving and d-DNNF compilation, recent advances in distributed model counting, such as \dcountAntom~\cite{DBLP:conf/cluster/BurchardSB16} and the more recent \discount~\cite{DBLP:conf/kr/LagniezL25}, provide a theoretical foundation for distributed {KC}.
These solvers typically rely on a \emph{Cube-and-Conquer} strategy, decomposing the search space into independent subproblems (cubes) distributed across a cluster.

While work-stealing approaches, such as \dmc~\cite{DBLP:conf/ijcai/LagniezMS18}, have proven effective for counting, extending them to compilation presents a unique challenge. 
In $\#\mathsf{SAT}$, workers only need to exchange scalar counts. 
In KC, however, workers must share and merge \emph{partial circuits}. 
Managing these circuit fragments across a distributed network introduces prohibitive communication overhead. 
Consequently, the Cube-and-Conquer paradigm is better suited for compilation, as it allows for a more structured reconstruction of the global target circuit.

Furthermore, unlike model counting where the process terminates once a value is returned, the output of compilation is a persistent artifact intended for subsequent querying. 
In this paper, we bridge the gap between distributed generation and distributed utilization. 
We introduce \dkc{}, the first distributed knowledge compiler based on a Cube-and-Conquer architecture, and \dreasoner{}, a distributed reasoner designed to evaluate compiled d-DNNF fragments across a cluster. 
Our reasoner supports distributed model counting, direct access, and uniform sampling, including queries under partial assignments (conditioning).

The remainder of this paper is organized as follows: Section~\ref{formal} provides the necessary formal background on the KC map and {d-DNNF}; Section~\ref{pkc} details the architectural design of the distributed compiler \dkc{}; Section~\ref{preasoner} describes the \dreasoner{} query engine; Section~\ref{empirical} presents an empirical evaluation on industrial benchmarks; and Section~\ref{conclusion} concludes with a summary and future perspectives.

\section{Preliminaries~\label{formal}}
\subsection{Propositional Logic and Notations}

We assume a standard propositional language $\mathcal{L}_{\mathcal{P}}$ built from a finite set of atoms $\mathcal{P}$ and the usual logical connectives ($\land, \lor, \to, \leftrightarrow, \neg$). 
This language $\mathcal{L}_{\mathcal{P}}$ follows classical semantics. 
For any formula $\Sigma \in \mathcal{L}_{\mathcal{P}}$, the set of propositional variables occurring in it is denoted by $\mathit{Var}(\Sigma)$. 
Given a finite variable set $X$, we denote by ${\{0, 1\}}^{X}$ the set of all possible Boolean assignments over $X$. 
Each formula $\Sigma \in \mathcal{L}_{\mathcal{P}}$ induces a Boolean function over its variables, mapping each assignment in $\{{0, 1\}}^{\mathit{Var}(\Sigma)}$ to a truth value in $\{0, 1\}$.

Assignments that satisfy $\Sigma$ (mapping it to 1) are called \emph{models} or \emph{satisfying assignments}, and the set of all such models is denoted by $\mathcal{M}(\Sigma)$. 
Formulas $\Sigma_1$ and $\Sigma_2$ are logically equivalent, denoted $\Sigma_1 \equiv \Sigma_2$, if and only if $\mathcal{M}(\Sigma_1) = \mathcal{M}(\Sigma_2)$. 
Logical entailment is defined as $\Sigma_1 \models \Sigma_2$ if $\mathcal{M}(\Sigma_1) \subseteq \mathcal{M}(\Sigma_2)$. 
The symbols $\bot$ and $\top$ represent the unsatisfiable and the tautological formulas, respectively.

A \emph{literal} is a variable $x$ or its negation $\neg x$. 
For any literal $\ell$, its underlying variable is denoted $\mathit{Var}(\ell)$ (where $\mathit{Var}(x) = \mathit{Var}(\neg x) = x$), and its complement is denoted ${\sim}\ell$ (with ${\sim}x = \neg x$ and ${\sim}\neg x = x$). 
The \emph{conditioning} of a formula $\Sigma$ by a literal $\ell$ (where $\ell = x$ or $\ell = \neg x$), denoted $\Sigma[\ell]$, is the formula resulting from replacing $x$ (resp. $\neg x$) with $\top$ and its complement with $\bot$. 
This is followed by a simplification process based on standard Boolean identities (e.g., $\top \lor \Gamma = \top$, $\top \land \Gamma = \Gamma$) until a fixed point is reached. 
This operation naturally extends to a set of literals $S = \{\ell_1, \ldots, \ell_m\}$ such that $\Sigma[S] = ((\Sigma[\ell_1]) \ldots)[\ell_m]$. 

An assignment $\mu$ is a set of literals interpreted conjunctively. 
We distinguish between \emph{total assignments}, which assign every variable in $\mathcal{P}$, and \emph{partial assignments}. 
A formula in \emph{Conjunctive Normal Form} (CNF) is a conjunction of clauses, where each clause is a disjunction of literals; for convenience, we view a CNF as a set of clauses and a clause as a set of literals.

\begin{example}\label{ex:running}
Let $\Psi = \{a \vee b, \neg a \vee \neg b, c \vee d \vee a, c \vee d \vee b\}$ be a CNF formula with $\mathit{Var}(\Psi) = \{a,b,c,d\}$.
\end{example}

\subsection{d-DNNF Representation}

The Deterministic Decomposable Negation Normal Form, or d-DNNF, is a specific type of Boolean circuit characterized by a single output node acting as its root. 
Structurally, a d-DNNF is represented as a rooted Directed Acyclic Graph (DAG), denoted $\langle V, E \rangle$. 
In this graph, leaf nodes consist of literals or Boolean constants ($\bot, \top$), while internal nodes are either decomposable $\land$ gates or deterministic $\vee$ gates. 
A gate $N = \wedge(N_1, \ldots, N_k)$ is \emph{decomposable} if the variables appearing in the sub-circuits rooted at $N_i$ and $N_j$ are disjoint for all $i \neq j$. 
An OR gate $N = \vee(N_1, \ldots, N_k)$ is \emph{deterministic} if its children $N_i$ and $N_j$ are pairwise inconsistent for any $i \neq j$. 
The size of a \dDNNF{} $\Sigma = \langle V,E \rangle$, denoted $|\Sigma|$, corresponds to the number of its edges $|E|$. 
As a language, \dDNNF{} is universal, meaning it is capable of representing any propositional theory~\cite{Darwiche02a}.

State-of-the-art \dDNNF{} compilers, including \ctwod~\cite{Darwiche04}, \dSharp~\cite{MuiseMBH12}, \dfour~\cite{LagniezM17}, and \sharpSAT~\cite{KieselE23}, typically generate \emph{decision-DNNF} representations. 
A decision-DNNF is a restricted form of d-DNNF where deterministic $\vee$ gates are replaced by decision gates of the form $N = \mathit{ite}(x, N_1, N_2)$. 
Here, $\mathit{ite}$ represents the ``{\tt if-then-else}'' connective, and the decision variable $x$ must not appear in the sub-circuits $N_1$ or $N_2$. 
Also known as decomposable decision graphs~\cite{FargierM06}, decision-DNNFs can be transformed into d-DNNFs in linear time. 
By expanding a decision node $N = \mathit{ite}(x, N_1, N_2)$ into the fragment $(\neg x \wedge N_1) \vee (x \wedge N_2)$, the properties of decomposability (since $x \notin \mathit{Var}(N_i)$) and determinism (due to the mutual exclusivity of $x$ and $\neg x$) are preserved.

\begin{example}[Example~\ref{ex:running} cont’ed]
For the \cnf{} formula $\Psi$ from Example~\ref{ex:running}, an equivalent \dDNNF{} representation is $\Sigma = ((\neg a \wedge b) \vee (a \wedge \neg b)) \wedge (c \vee (\neg c \wedge d))$.
\end{example}

The d-DNNF language is a highly effective representation for automated reasoning, as it supports a wide range of queries and transformations~--~such as satisfiability checking and conditioning~--~in polynomial time. 
Importantly, tasks like direct access~\cite{CarmeliTGKR23,DBLP:journals/tods/BringmannCM25}, uniform sampling~\cite{SharmaGRM18}, and model counting~\cite{Darwiche02a} can be performed efficiently when the input formula is compiled into a \dDNNF{}.

\subsection{Queries}

Beyond the theoretical elegance of d-DNNF, its primary value lies in its support for efficient reasoning. 
In real-world scenarios (ranging from interactive product configuration~\cite{DBLP:journals/corr/abs-1007-1024,DBLP:conf/se/SundermannHNBYT24,DBLP:conf/splc/OhGB19,DBLP:conf/icst/PlazarAPDC19} to probabilistic reasoning~\cite{DBLP:journals/ai/ChaviraD08,DBLP:conf/ecai/BartKLM16} and system verification~\cite{DBLP:conf/icse/BalutaCMS21,DBLP:conf/somet/HeradioPFBCCH16}) practitioners require more than just a compact representation; they need to extract specific insights efficiently. 
Core tasks such as checking consistency, estimating probabilities, or generating diverse scenarios rely on four fundamental queries: \emph{Satisfiability}, \emph{Model Counting}, \emph{Direct Access}, and \emph{Uniform Sampling}. 
While these tasks are generally intractable ($\mathsf{NP}$-complete or $\#\mathsf{P}$-complete) on raw \cnf{} formulas, they become polynomial-time operations when the formula is compiled into a \dDNNF{}.

\paragraph{Satisfiability.} 
The \emph{Satisfiability} (SAT) check is the most fundamental query: it determines whether there exists at least one assignment $\mu$ such that $\mu \models \Psi$ (i.e., if $\norm{\Psi} > 0$). 
While SAT is the canonical $\mathsf{NP}$-complete problem, it is trivial on a {d-DNNF}. 
Due to the properties of decomposability and determinism, a \dDNNF{} circuit is satisfiable if and only if its root is not equivalent to the constant $\bot$. 
This check can be performed in constant time if the compiler prunes inconsistent branches during construction, or in linear time otherwise.

\paragraph{Model Counting.} The \emph{Model Counting} problem (a.k.a. $\#\mathsf{SAT}$) involves determining the total number of satisfying assignments for a given propositional formula $\Phi$, denoted by $\norm{\Phi}$. 
This task is the functional counterpart to the decision problem and is known to be $\#\mathsf{P}$-complete. 
While computationally intensive for general formulas, model counting becomes a linear-time operation on d-DNNF representations.
This is achieved by traversing the circuit bottom-up: AND gates return the product of their children's counts (due to decomposability), and OR gates return the sum of their children's counts (due to determinism).

\begin{example}[Ex.~\ref{ex:running} cont.]\label{ex:firstModelEnum}
    For the formula $\Psi$ from Example~\ref{ex:running}, $\norm{\Psi} = 6$ and the models are:  
    $\{a,\neg b,c,d\}$, $\{\neg a, b,c,d\}$, $\{a,\neg b,c, \neg d\}$, $\{\neg a, b,c, \neg d\}$, $\{a,\neg b,\neg c,d\}$, and $\{\neg a, b, \neg c,d\}$. 
\end{example}

\paragraph{Direct Access.} The \emph{direct access} task, originally introduced in the database literature~\cite{BaganDGO08}, requires returning the $k$-th model of a formula $\Psi$ according to a lexicographical order $\prec_{lex}$. 
If $k$ exceeds the total model count $\norm{\Psi}$, the query fails. 
In propositional logic, $\prec_{lex}$ is defined by fixing an ordering $\tau$ over the variables and treating assignments as binary words in ${\{0,1\}}^{\mathit{Var}(\Psi)}$.
We denote this ordered relation as $\prec_{lex}^{\tau}$, using $x <_{\tau} y$ to indicate that variable $x$ precedes $y$ in $\tau$.

\begin{example}[Ex.~\ref{ex:running} cont.]\label{ex:firstDirectAccess}
Using $\tau = (a,b,c,d)$ for the formula $\Psi$, the first model (word $0101$) is $\{\neg a, b, \neg c, d\}$, and the third model (word $0111$) is $\{\neg a, b, c, d\}$.
\end{example}

Direct access serves as a fundamental primitive for advanced tasks like uniform sampling without replacement~\cite{SharmaGRM18}.
While generally $\#\mathsf{P}$-hard, the query is tractable for d-DNNF because the language supports polynomial-time counting. 
The algorithm iteratively selects the next variable $x$ in $\tau$. Starting with an empty assignment $\sigma$, it checks if the count of the branch $\neg x$ (i.e., $\norm{\Psi[\sigma \cup \{\neg x\}]}$) is at least $k$. If so, $\neg x$ is added to $\sigma$; otherwise, $x$ is added, and $k$ is offset by the count of the $\neg x$ branch.

\begin{example}[Ex.~\ref{ex:firstDirectAccess} cont.]
To find the third model of $\Psi$, we start with $\sigma = \emptyset$. Since $\norm{\Psi[\neg a]} = 3 \geq 3$, we set $\sigma = \{\neg a\}$. 
Next, for variable $b$, we find $\norm{\Psi[\{\neg a, \neg b\}]} = 0$. 
Since $0 < 3$, we set $\sigma = \{\neg a, b\}$. 
For $c$, $\norm{\Psi[\{\neg a, b, \neg c\}]} = 1 < 3$, so we select $c$, resulting in $\sigma = \{\neg a, b, c\}$. Finally, variable $d$ is determined, yielding $\{\neg a, b, c, d\}$.
\end{example}

\paragraph{Uniform Sampling.} Uniform sampling involves a generator $\mathcal{G}$ that produces a sample from $\mathcal{M}(\Psi)$ such that every model has an equal probability $1/\norm{\Psi}$ of being selected. 
A standard technique for achieving seed-based reproducible uniform sampling involves determining the total model count $c$, generating a uniform random integer $k \in \{1, \dots, c\}$, and retrieving the corresponding model via a direct access query~\cite{DBLP:conf/jelia/LagniezL25}. 
In contrast, the approach proposed in~\cite{SharmaGRM18} tags the circuit to compute uniform samples more efficiently. 
However, such methods do not impose a strict order on models, meaning the output is inherently dependent on the specific circuit structure. 
Since compilers generally lack deterministic control over the resulting circuit topology, these structure-dependent methods make seed-based reproducibility difficult to achieve, even though they are typically faster to compute.

\paragraph{Inference under Conditioning.} A critical requirement for real-world reasoning is the ability to answer these queries under specific contexts. 
Given a set of evidence literals $E$, all aforementioned queries~--~satisfiability, model counting, direct access, and uniform sampling~--~can be performed on the conditioned formula $\Psi[E]$. 
Because d-DNNF is closed under conditioning, $\Psi[E]$ remains a d-DNNF, ensuring that inference remains polynomial even when restricted to a sub-space of the models.

\section{Distributed d-DNNF Compiler~\label{pkc}}
In this section, we present the architecture of \dkc{}, our distributed framework for knowledge compilation. 
Our approach builds upon the \emph{Cube-and-Conquer} paradigm effectively employed in distributed model counter \discount, but introduces fundamental structural changes to support the generation and persistent storage of circuit fragments, as well as an online query processing phase.

\subsection{The Cube-and-Conquer Paradigm}

The foundation of our approach is the decomposition of the initial formula $\Psi$ into a set of independent subproblems using a Lookahead-based partitioning strategy. 
Formally, given a propositional formula $\Psi$, the Cube-and-Conquer method identifies a set of \emph{cubes} (partial assignments) $\mathcal{T} = \{\tau_1, \tau_2, \dots, \tau_m\}$ such that:
\begin{equation}
    \Psi \equiv \bigvee_{\tau \in \mathcal{T}} (\Psi \wedge \tau)
\end{equation}
Crucially, these cubes are generated to be pairwise disjoint ($\tau_i \wedge \tau_j \models \bot$ for $i \neq j$) and to cover the entire solution space of $\Psi$. 
This disjointness is particularly advantageous for d-DNNF compilation, as it naturally satisfies the \emph{determinism} property at the root level of the global circuit.

In standard distributed counting, a Master dispatches cubes to a worker set $\mathcal{W}$ of size $N = |\mathcal{W}|$. 
Each $W_j \in \mathcal{W}$ simplifies a cube $\tau$ to $\Psi_{|\tau}$, returns its count, and discards the task. 
The Master then aggregates the global sum $\sum_{\tau \in \mathcal{T}} \norm{\Psi \wedge \tau}$.

\subsection{From Transient Counting to Persistent Compilation}

Although the ``count-and-forget'' approach is sufficient for $\#\mathsf{SAT}$, Knowledge Compilation requires the preservation of the structural information computed by the workers. 
To achieve this, we modify the worker lifecycle and the Master-Worker protocol as follows.

\paragraph{Distributed Compilation.} 
Instead of a model counter, each worker is equipped with a sequential d-DNNF compiler (\dfour{} in our experiments). 
Upon receiving a cube $\tau$, the worker compiles the conditioned formula $\Psi \wedge \tau$ into a d-DNNF fragment $\Sigma_{\tau}$.
Unlike the counting approach, the worker does \emph{not} return the explicit circuit $\Sigma_{\tau}$ to the Master at the end.
Transmitting large circuits over the network would create a significant bottleneck and likely overwhelm the Master's memory. 
Instead, the worker appends $\Sigma_{\tau}$ to a persistent, \textbf{ordered list} $L_j$ stored in local memory.
\begin{equation}
    L_j = \langle \Sigma_{1}, \Sigma_{2}, \dots, \Sigma_{k} \rangle
\end{equation}
where $\Sigma_{i}$ denotes the $i$-th d-DNNF fragment compiled by worker $W_j$.
The worker then acknowledges the completion of the task to the Master, but the circuit itself remains distributed.
Maintaining a strict order is crucial for the subsequent reasoning phase, as it allows the worker to target specific circuit fragments by their index when performing sampling or direct access queries.

\paragraph{The Global Virtual Circuit.}
From a logical perspective, the global d-DNNF $\Delta$ representing $\Psi$ effectively exists as a virtual structure distributed across the cluster. 
It can be viewed as a large deterministic OR gate located at the Master, with edges connecting to the roots of the local d-DNNF fragments stored in the lists of the workers:
\begin{equation}
    \Delta = \bigvee_{j=1}^{N} \left( \bigvee_{\Sigma \in L_j} \Sigma \right)
\end{equation}
The correctness of this virtual reconstruction follows directly from the Cube-and-Conquer partition: the cubes cover the whole solution space, and their pairwise inconsistency ensures determinism at the top OR gate.
By keeping the fragments $\Sigma$ local to the workers, \dkc{} overcomes the memory wall that limits centralized compilers, allowing for the representation of formulas significantly larger than the RAM capacity of a single machine.
This also motivates the distributed reasoning phase: collecting all fragments on one node would often be impractical, as the resulting circuit can exceed the memory available on any single machine.

\subsection{Distributed Query Processing}

The transition from offline compilation to online reasoning marks the second fundamental shift in our architecture. 
Unlike the standard distributed counting workflow, which terminates immediately after computing a global sum, \dkc{} initializes a persistent query engine, \dreasoner{}, once the compilation phase concludes.

In this phase, the Master node evolves from a \emph{Task Dispatcher} into a \emph{Query Coordinator}. 
The workflow for handling a query $\mathcal{Q}$ (whether for satisfiability, model counting, direct access, or uniform sampling) proceeds as follows:

\begin{enumerate}
    \item \textbf{Query Broadcast:} The Master receives a query request $\mathcal{Q}$, potentially accompanied by a set of conditioning literals $\gamma$ (evidence). The Master broadcasts $\mathcal{Q}$ and $\gamma$ to all participating workers.
    
    \item \textbf{Local Evaluation:} Each worker $W_j$ executes the query against its local knowledge base. Specifically, the worker iterates through its persistent, ordered list $L_j$ of d-DNNF fragments. For each fragment $\Sigma \in L_j$, the worker computes the query result locally (e.g., counting models consistent with $\gamma$). The specific algorithmic procedures for these local evaluations are detailed in the next section.
    
    \item \textbf{Global Aggregation:} Workers transmit their local results back to the Master. The Master then aggregates these partial responses to derive the final global answer. 
    For instance, in a model counting query, this involves summing the scalar counts returned by each worker ($C = \sum c_j$); for direct access or sampling, it involves identifying the specific worker holding the target model.
\end{enumerate}

This architecture guarantees that traversing the complex d-DNNF structures remains fully distributed across the cluster, while the Master is responsible only for lightweight coordination and scalar aggregation.

\subsection{The Compilation Algorithm}

\begin{algorithm}[t]
\LinesNumbered{}
\caption{\dkc{} Compilation Control Loop}\label{alg:dkc_compilation}
\KwData{
    \begin{itemize}
        \item[-] $\Sigma$: a CNF formula;
        \item[-] $nbCubes$: granularity factor (integer); 
        \item[-] $\mathcal{W}$: the set of workers.
    \end{itemize}}
\KwResult{A \dDNNF{} ready for querying.}

\BlankLine{}
\tcp{1. Preprocessing and Broadcast}
$\Sigma \gets \mathtt{preprocessing}(\Sigma)$\;
$\mathtt{broadcast}(\Sigma, \mathcal{W})$\;

\BlankLine{}
\tcp{2. Search Space Partitioning}
$\mathcal{C} \gets \mathtt{generateCubes}(\Sigma, nbCubes \times |\mathcal{W}|, \mathcal{W})$\;

\BlankLine{}
\tcp{3. Distributed Compilation Loop}
$idle[w] \gets 1$ for each $w \in \mathcal{W}$\;
$\mathtt{setWorkerCompilationMode}(\mathcal{W})$\;
\BlankLine{}
\While{$\mathcal{C} \neq \emptyset$}{
    \If{$\exists\, w \in \mathcal{W}$ s.t. $idle[w] = 1$}{
        $\tau \gets \mathtt{pop}(\mathcal{C})$\;
        $\mathtt{sendCubeToCompile}(\tau, w)$\;
        $idle[w] \gets 0$\;
    }
    \tcp{Poll for task completion}
    \While{$\exists\, w \in \mathcal{W}$ s.t. $\mathtt{free}(w)$ and $idle[w] = 0$}{
        $idle[w] \gets 1$\;
    }
}

\BlankLine{}
\tcp{4. Barrier Synchronization}
\While{$\exists\, w \in \mathcal{W}$ s.t. $idle[w] = 0$}{
     \lIf{$\mathtt{free}(w)$}{$idle[w] \gets 1$}
}

\BlankLine{}
\tcp{5. Transition to Reasoning}
$\mathtt{dreasoner}(\mathcal{W})$\; 

\BlankLine{}
\tcp{6. Termination}
\lFor{$w \in \mathcal{W}$}{$\mathtt{stop}(w)$}

\BlankLine{}
\Return{}
\end{algorithm}

The control logic of \dkc{}, outlined in Algorithm~\ref{alg:dkc_compilation}, is directly derived from the \discount{} framework introduced in~\cite{DBLP:conf/kr/LagniezL25}. 
We explicitly retain the core infrastructure of that approach, including the preprocessing pipeline, the formula broadcast mechanism, and the lookahead-based cube generation strategy, which effectively partitions the search space. 
Consequently, the Master's control loop remains structurally similar to the original counting architecture. 
The fundamental difference lies in the nature of the distributed task: whereas \discount{} workers reduce sub-problems to scalar counts, \dkc{} workers now act as compilers, generating and persisting d-DNNF fragments for future reasoning.

\paragraph{Preprocessing and Partitioning (Lines 1--3).}
The process begins with a preprocessing step on the Master node (line 1). 
Standard simplification techniques (e.g., unit propagation, vivification, \ldots) are applied to the input formula $\Sigma$ to reduce its size before transmission. 
The simplified formula is then broadcast to all worker nodes $\mathcal{W}$ (line 2). 
Subsequently, the Master generates a set of disjoint cubes $\mathcal{C}$ using the heuristic inherited from \discount{} (line 3). 
The number of cubes is determined by a scaling factor ($nbCubes$) relative to the cluster size $|\mathcal{W}|$, ensuring enough granularity to smooth out runtime variations among workers.

\paragraph{Dynamic Dispatch Loop (Lines 3--12).}
The Master utilizes a dynamic scheduling strategy to manage the workload:
\begin{enumerate}
    \item \textbf{Assignment:} As long as there are cubes remaining in $\mathcal{C}$ and idle workers are available, the Master pops a cube $\tau$ and dispatches the task ($\mathtt{sendCubeToCompile}$) to the target worker (lines 7--10).
    \item \textbf{Completion Check:} The Master continuously polls for completion signals ($\mathtt{free(w)}$). 
    When a worker $w$ finishes compiling its assigned fragment (and appends it to its local list $L_w$), it notifies the Master, which then marks $w$ as available to receive a new cube (lines 11--12).
\end{enumerate}

\paragraph{Finalization and Transition (Lines 13--16).}
Once the dispatch queue $\mathcal{C}$ is empty, the Master enters a synchronization barrier, strictly waiting for all currently active workers to complete their final tasks and return to an idle state (lines 13--14). 
Crucially, unlike counting algorithms that terminate immediately after this point, \dkc{} preserves the distributed state and transitions into the \emph{Online Phase} (line 15). 
The function $\mathtt{dreasoner}(\mathcal{W})$ initializes the Query Coordinator loop (described in Section~\ref{preasoner}), effectively blocking the main execution flow to accept and process queries against the distributed circuit. 
Only when the reasoning session is explicitly ended by an external signal does the control flow return to terminate the workers (line 16).

\section{Distributed d-DNNF Reasoner~\label{preasoner}}
This section describes how the \dreasoner{} engine handles query processing.
Unlike traditional counting solvers that terminate immediately after reporting a scalar, \dkc{} transitions into a persistent serving state upon the completion of Algorithm~\ref{alg:dkc_compilation}.
In this state, the global d-DNNF $\Delta$ is implicitly represented by the union of disjoint fragments distributed across the worker nodes.
Consequently, the Master node shifts its role from a \emph{Task Dispatcher} to a \emph{Query Coordinator}, executing an event loop to process incoming requests.
We categorize the supported operations into two classes: \emph{Aggregation Queries} (satisfiability and model counting), which summarize properties of the entire solution space, and \emph{Member Queries} (uniform sampling and direct access), which retrieve specific model instances.

\subsection{Aggregation Queries}

Aggregation queries compute global properties over the distributed d-DNNF $\Delta$. Exploiting the decomposability and determinism of the fragments, these operations naturally map to a parallel reduction pattern.

\paragraph{Satisfiability.}
To check satisfiability under a partial assignment $\gamma$, the Master broadcasts $\gamma$ to all workers.
Each worker $W_j$ scans its local fragments $L_j$; if any $\Sigma \in L_j$ is consistent with $\gamma$, the worker reports \texttt{true}.
Formally, the global satisfiability is the disjunction of local consistency checks:
\begin{equation}
\mathtt{SAT}(\Delta \wedge \gamma) = \bigvee_{j=1}^{N} \left( \bigvee_{\Sigma \in L_j} \mathtt{SAT}(\Sigma \wedge \gamma) \right) 
\end{equation}

\paragraph{Model Counting.}
Analogously, for model counting, the Master broadcasts the request (potentially with weights or conditioning $\gamma$).
Each worker sums the model counts of its resident fragments, returning a scalar partial sum to the Master, which computes the global total:
\begin{equation}
    \mathtt{\#SAT}(\Delta \wedge \gamma) = \sum_{j=1}^{N} \sum_{\Sigma \in L_j} \mathtt{\#SAT}(\Sigma \wedge \gamma)
\end{equation}
Crucially, since the fragments are persisted in worker memory, computing $\mathtt{\#SAT}(\Sigma \wedge \gamma)$ reduces to a graph traversal linear in the size of the local fragments.
This avoids the latency of disk I/O or recompilation, ensuring high throughput even under arbitrary conditioning.

\subsection{Member Queries}

Member queries involve generating full assignments (models) from the solution space. 
These operations are \emph{count-based}, relying on the model counting primitive to navigate the decision tree structure.

\paragraph{Direct Access.}
The Direct Access query retrieves the $k$-th model of the formula according to a fixed global variable ordering $\prec_{lex}$.
Since the \decDNNF{} is distributed, the Master must orchestrate a global search variable by variable.
Let $x$ be the current variable in $\prec_{lex}$, $\gamma$ the current partial assignment, and $k$ the target index.
At each step, the Master tentatively branches on $\neg x$ by querying the global count $c$ of the subspace consistent with $\gamma \land \neg x$.
If $c \ge k$, the target model lies in the $\neg x$ branch; the Master updates $\gamma \gets \gamma \cup \{\neg x\}$.
Otherwise, the model resides in the $x$ branch; the Master updates $\gamma \gets \gamma \cup \{x\}$ and adjusts the target index $k \gets k - c$.
This procedure repeats for all variables until a complete model is constructed.

\paragraph{Uniform Sampling.}
Uniform sampling generates a model such that every solution has an equal probability $1/N$ of being selected.
Our architecture implements a two-stage routing protocol that leverages the disjointness of the cubes to minimize network overhead.
First, the Master issues a global \emph{Model Counting} query to determine the total number of solutions $C$, and generates a set of random indices $\mathcal{S} = \{s_1, \dots, s_m\}$ where each $s \in [1, C]$.
Instead of iterating through variables as in Direct Access, the Master directly identifies the worker $W_{t}$ that ``owns'' the $s$-th solution using the prefix sums of the worker counts $\{c_j\}$.
Specifically, $W_{t}$ is the worker satisfying $\sum_{j < t} c_j < s \le \sum_{j \le t} c_j$.
The query is then unicast exclusively to $W_{t}$ with the local index $s' = s - \sum_{j < t} c_j$, instructing the worker to traverse $L_{t}$ and retrieve the $s'$-th model.

Crucially, when generating a batch of samples under the same evidence $\gamma$, the model counts for every node in the circuit are cached during the initial global counting phase.
Consequently, subsequent sampling queries reuse these values to navigate the circuit branches, ensuring the expensive counting operation is amortized across the entire batch.

\begin{algorithm}[t!]
  \LinesNumbered{}
  \caption{\dreasoner{} Master Control Logic}\label{alg:dreasoner_logic}
  \KwData{
    \begin{itemize}
    \item $\mathcal{W}$: the set of workers;
    \item $\Delta$: global \decDNNF.
    \end{itemize}}
  
  \SetKwFunction{FProcess}{ProcessQuery}
  \SetKwProg{Fn}{Function}{:}{}
  
  \BlankLine{}
  \tcp{Main Event Loop}
  \While{\textup{true}}{
      $\mathcal{Q} \gets \mathtt{WaitForQuery()}$\;
      \lIf{$\mathcal{Q} = \mathtt{terminate}$}{\text{break}}
      \BlankLine{}

      $result \gets $ \FProcess{$\mathcal{Q}, \mathcal{W}$}\;
      $\mathtt{Display}(result)$\;
  }
  \BlankLine{}
  
  \tcp{Query Processing Function}
  \Fn{\FProcess{$\mathcal{Q}, \mathcal{W}$}}{
    $\gamma \gets \mathcal{Q}.\mathtt{evidence}$\;
    
    \Switch{$\mathcal{Q}.\mathtt{type}$}{
      \Case{\textup{Satisfiability}}{
        \Return{} $\mathtt{SAT}(\Delta \land \gamma)$\;
      }
      \Case{\textup{Model Counting}}{
        \Return{} $\mathtt{\#SAT}(\Delta \land \gamma)$\;
      }
      \Case{\textup{Direct Access}}{
        $\mu \gets \gamma$\;
        $k \gets \mathcal{Q}.k$\;

        \BlankLine{}
        \tcp{Binary search}
        \ForEach{$x \in \mathtt{Vars}(\Psi)$ ordered by $\prec_{lex}$}{
          \lIf{$x \in \mathtt{Vars}(\mu)$}{\text{continue}}
          \BlankLine{}

          $c \gets \mathtt{\#SAT}(\Delta \land \mu \land \neg x)$\;
          \lIf{$c \ge k$}{
            $\mu \gets \mu \cup \{\neg x\}$
          }
          \Else{
            $\mu \gets \mu \cup \{x\}$\;
            $k \gets k - c$\;
          }
        }
        \Return{} $\mu$\;
      }
      \Case{\textup{Uniform Sampling}}{
        $C, \{c_j\} \gets \mathtt{\#SAT}(\Delta \land \gamma)$\;
        $\mathcal{S} \gets \mathtt{GenRandomIndices}(C, \mathcal{Q}.size)$\;
        $\Gamma  \gets \emptyset$\;
        \ForEach{$s \in \mathcal{S}$}{
          $w_{t} \gets \mathtt{SelectWorker}(s, \{c_j\})$\;
          $s' \gets s - \sum_{j < t} c_j$\;
          $\Gamma \cup \mathtt{Sample}(w_{t}, s', \gamma)$\;
        }
        \Return{} $\Gamma$\;
      }
    }
  }
\end{algorithm}

\subsection{Reasoning Algorithm}

The control logic of the \dreasoner{}, outlined in Algorithm~\ref{alg:dreasoner_logic}, operates as a continuous, event-driven service.
Unlike the linear compilation phase, the reasoning engine executes an infinite loop to process a stream of incoming queries until an explicit termination signal is received.

The Master blocks on the reception of client requests (Lines 1--5).
Upon receiving a query $\mathcal{Q}$, it validates the active state and invokes the \texttt{ProcessQuery} dispatcher, which coordinates the distributed workers $\mathcal{W}$ based on the query type and evidence $\gamma$.
Finally, upon completion of the reasoning task, the computed result is formatted and emitted via the \texttt{Display} function.

For \emph{satisfiability} (Lines 9--10) and \emph{model counting} (Lines 11--12), the algorithm follows a map-reduce pattern.
The Master broadcasts the request to all workers and aggregates their partial results (via disjunction or summation) to return the global answer.

For \emph{direct access} (Lines 13--23), the algorithm constructs the model by determining the assignment of each variable $v$ according to the global order $\prec_{lex}$.
At each step, it broadcasts a query to count the models consistent with the negative branch ($\gamma \land \neg x$) (Line 18).
Let $c$ be this count; if $k \le c$, the target model resides in the negative branch, so the algorithm commits to $\neg x$ (Line 19).
Otherwise, it deduces the model implies $x$, updates the target rank $k \gets k - c$, and proceeds to the next variable (Lines 21--22).

For \emph{uniform sampling} (Lines 24--32), the algorithm employs a routing-based strategy.
After retrieving the solution distribution $\{c_j\}$ and the global total $C$ (Line 25), the Master maps a batch of random global indices $\mathcal{S}$ to specific workers (Line 26).
It then unicasts retrieval requests to the identified worker (Line 29), explicitly querying for the model at the local index $s'$ corresponding to the target global position within that worker's fragment (Lines 30--31).

\section{Experiments~\label{empirical}}
\subsection{Implementation and Environment}
Our distributed architecture, \dkc{}, is implemented in \texttt{C++} as an extension of the open-source \discount{} framework,\footnote{\url{https://zenodo.org/records/16536062}} adhering to the configuration guidelines established in~\cite{DBLP:conf/kr/LagniezL25}. 
The software used in the experiments, along with the corresponding logs, is available at https://zenodo.org/records/19936190.
The system integrates several state-of-the-art components to optimize performance. 
For Boolean constraint propagation and solving, we employ CaDiCaL~\cite{DBLP:conf/cav/BiereFFFFP24} as the underlying SAT solver, maintaining a single incremental solver instance to minimize overhead across repeated calls. 
Structural decomposition is handled by \texttt{FlowCutter}~\cite{DBLP:journals/jea/HamannS18}, specifically the implementation from the PACE 2017 challenge,\footnote{\url{https://github.com/kit-algo/flow-cutter-pace17.git}} with a strict time budget of 10 seconds allocated for computing the tree decomposition. 

Prior to compilation, the formula undergoes simplification via the \texttt{B+E} preprocessor~\cite{DBLP:journals/ai/LagniezLM20}.\footnote{\url{http://www.cril.univ-artois.fr/kc/bpe2.html}} 
We utilize the \texttt{equiv} strategy (combining vivification, backbone extraction, and occurrence elimination~\cite{DBLP:journals/jar/LagniezM17}) limited to a 5-second budget. 
Finally, for the worker nodes, we integrate the \dfour{} compiler~\cite{DBLP:conf/ijcai/LagniezM17}.\footnote{\url{https://github.com/crillab/d4v2}} 
Crucially, \dfour{} is operated in \emph{incremental mode}, allowing it to preserve its cache between different compilation tasks (cubes), thereby avoiding redundant computations during the local compilation phase.

All experiments were conducted on a cluster comprising 32 nodes, each equipped with an Intel\textsuperscript{\textregistered} Xeon\textsuperscript{\textregistered} {E5}-2643 v4 CPU (3.30\,GHz) and running Rocky Linux 9.5 (kernel 5.14).
The nodes are interconnected via a 1\,GiB/s Ethernet network.
All runs were fully distributed, with one MPI process allocated per physical core.
The software environment was compiled using GCC 11.5 and Open~MPI 5.1.0a1.

Our evaluation is divided into two parts.
First, we assess the performance of the compilation process using benchmarks from the most recent Model Counting Competition (MC 2025), as detailed in Section~\ref{subsec:evalKC}.
Second, we evaluate the efficiency of the reasoner in Section~\ref{subsec:evalQuery}.
For this latter task, we focus on a subset of benchmarks widely used for querying assessment, specifically selecting instances that are tractable for compilation to isolate the reasoning performance.

\subsection{Evaluating the Compilation Process\label{subsec:evalKC}}

\begin{figure}[t!]
    \includegraphics[width=0.472\textwidth]{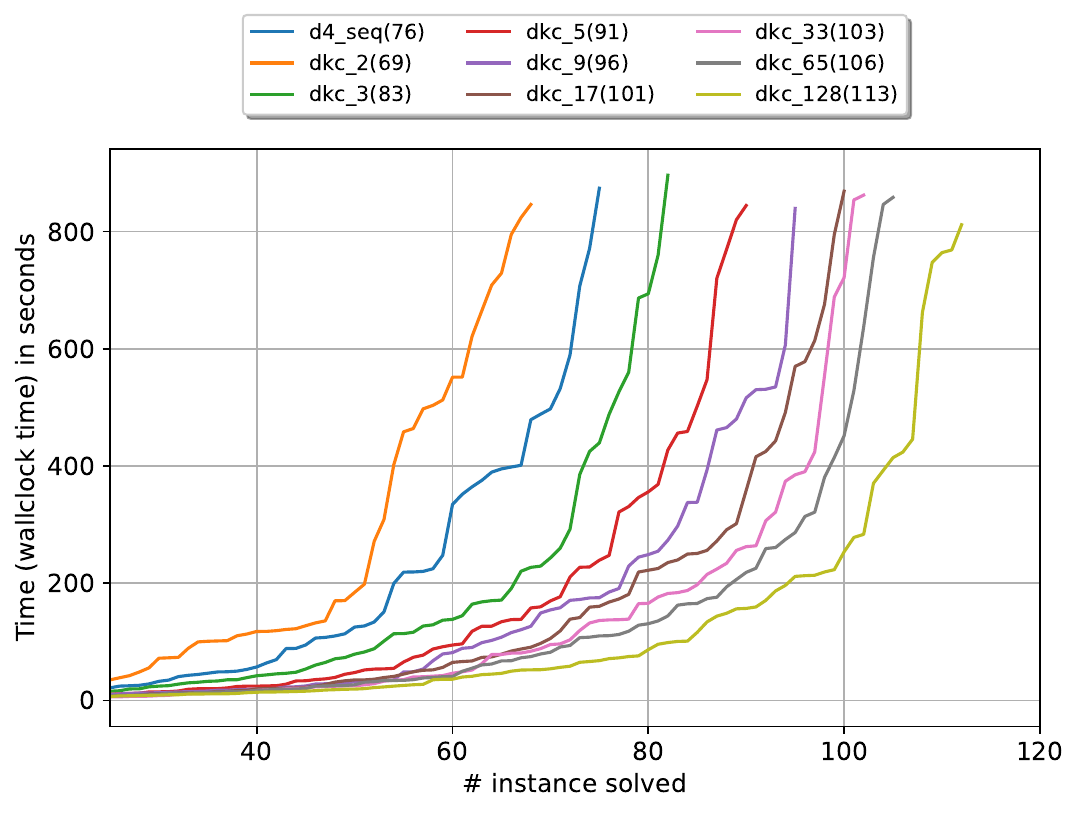}
    \caption{Performance comparison of the model counters \dfour{} (\dfour\_seq) and various versions of \dkc{} for different numbers of cores. The plot reports the number of instances solved as a function of time. For each solver, the total number of instances solved is indicated in parentheses in the legend.\label{fig:cactusTimeKC}}
\end{figure}

To evaluate the efficiency of \dkc{}, we utilized benchmarks from the most recent Model Counting Competition (MC 2025).\footnote{\url{https://nextcloud.liu.se/s/MfLJWiDfYzGjXs8}}
%
%
A wall-clock time limit of 900 seconds and a memory limit of 32\,GiB were enforced for each run.
Figure~\ref{fig:cactusTimeKC} presents a cactus plot illustrating the number of instances solved ($x$-axis) within a given wall-clock time ($y$-axis) by \dfour{} and various \dkc{} configurations.
To assess scalability, we varied the total core count from 2 to 128. Since \dkc{} designates a dedicated Master node, these configurations correspond to $1, 2, 4, \dots, 127$ active workers.

We first observe that the sequential solver \dfour{}, with 76 solved instances, outperforms \dkc{} when restricted to a single worker, which solves only 69.
This performance gap indicates that the overhead of the Cube-and-Conquer strategy, specifically the management of cubes and network communication, is detrimental when no parallelism is available to offset the cost.
However, this overhead is quickly amortized; with just two workers, \dkc{} solves 83 instances, surpassing the sequential baseline.
Increasing the worker count significantly boosts throughput, enabling the solution of progressively harder instances within the timeout.
Specifically, the number of solved instances scales robustly: 91 with 4 workers, 96 with 8 workers, and continuing to climb steadily to 113 solved instances with 127 workers.

\begin{figure}[t!]
    \includegraphics[width=0.472\textwidth]{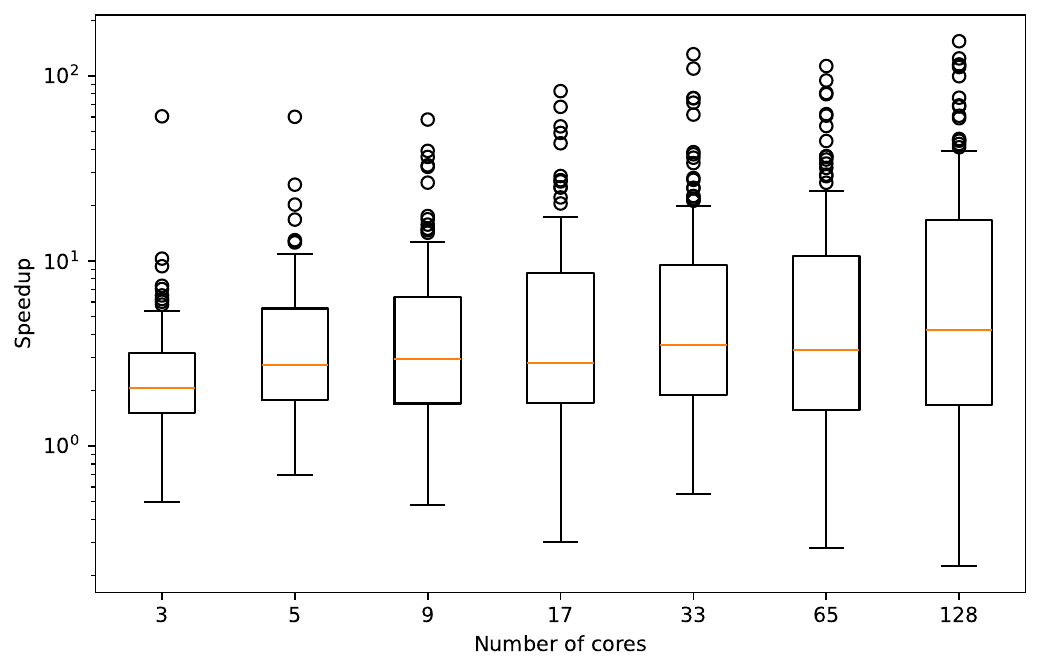}
    \caption{Distribution of speedups achieved by \dkc{} across varying core counts, normalized against the minimal distributed configuration (1 Master + 1 Worker).\label{fig:boxplot}}
\end{figure}

To assess the scalability of \dkc{}, Figure~\ref{fig:boxplot} reports the distribution of speedups as the cluster size expands from 3 to 128 cores.
The $y$-axis (log scale) represents the speedup factor relative to the baseline configuration of 2 cores (1 Master + 1 Worker).
For a given instance solved in $t_n$ seconds using $n$ cores, the speedup is calculated as $\frac{\min(t_{\text{2}}, 900)}{t_n}$, where $t_{\text{2}}$ is the wall-clock time of the baseline run.
Crucially, since $t_{\text{2}}$ is capped at the 900-second timeout, these values represent a \emph{conservative lower bound} on the actual speedup for any instance that times out in the baseline configuration.

The results demonstrate robust scalability.
We observe a consistent upward trend in the median speedup (orange line) as the core count increases, confirming that \dkc{} effectively translates additional worker nodes into reduced runtime.
Notably, the distribution exhibits significant positive skew; the high-performing outliers indicate that for computationally intensive instances, the framework achieves near-linear speedups.
This suggests that the Cube-and-Conquer strategy successfully amortizes the communication and synchronization overhead, particularly when the workload is sufficient to saturate the distributed workers.

\begin{figure}[t!]
    \includegraphics[width=0.472\textwidth]{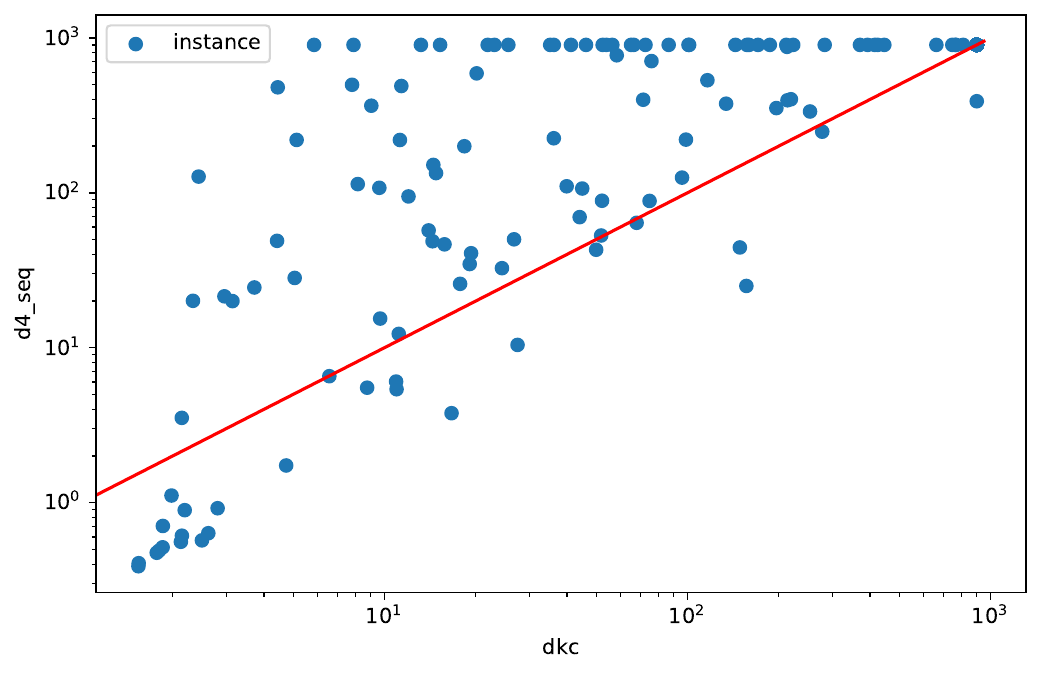}
    \caption{Scatter plot comparing the solving times of \dkc{} with 128 cores ($x$-axis) versus the sequential baseline \dfour{} ($y$-axis). 
    Points above the diagonal indicate instances where the distributed approach outperforms the sequential one.\label{fig:scatterD4vsDkc}}
\end{figure}

To conclude the evaluation of the compilation process, Figure~\ref{fig:scatterD4vsDkc} presents a scatter plot comparing the runtime of the sequential baseline \dfour{} against the fully distributed \dkc{} configuration (128 cores).
Each data point represents a single benchmark instance, with its coordinates $(x, y)$ corresponding to the solving time of \dkc{} and \dfour{}, respectively.
The results clearly demonstrate a threshold effect: for harder instances, specifically those requiring more than 100 seconds sequentially, the distributed approach consistently pays off, yielding significant speedups.
However, the presence of outliers below the diagonal (on the right) reveals that for certain instances, the Cube-and-Conquer decomposition may be detrimental.
In these cases, the overhead of partitioning the search space likely disrupts the variable ordering heuristics that allow the sequential solver to find a compact \decDNNF, resulting in a net performance loss despite the parallelism.

\subsection{Evaluating the Query Process\label{subsec:evalQuery}}

Our evaluation utilizes benchmarks from~\cite{DBLP:conf/jelia/LagniezL25}.\footnote{\url{https://zenodo.org/records/15837216}}
These instances are particularly suitable for evaluating \dreasoner{} as they represent concrete applications where compilation into \decDNNF{} is mandatory.
Due to computational constraints, we did not consider the full suite of 1,425 benchmarks; instead, we randomly selected a representative subset of 200 instances.
We impose a global wall-clock time limit of 1800 seconds and a memory limit of 32\,GiB per run.
To strictly evaluate each phase of our architecture, we separately measure the time required for distributed compilation ($t_c$) and the time required to answer the query batch ($t_q$).
To prevent a single phase from monopolizing the resource budget, we enforce a local timeout of 900 seconds for each; a run is considered a timeout if either $t_c > 900$\,s or $t_q > 900$\,s.

To evaluate the \dreasoner{} engine, we generated a synthetic workload designed to mimic real-world usage patterns.
The workload construction begins by computing a single satisfying assignment (seed model) $\omega$ for each instance using the \texttt{kissat} solver.
We generate a batch of $n=1,000$ queries per instance, distributed according to the following profile: $50\%$ Satisfiability, $30\%$ Model Counting, $10\%$ Uniform Sampling, and $10\%$ Direct Access.

The complexity of the queries is regulated by varying the size of the evidence set $\gamma$.
For each query in the sequence, the number of conditioning literals $|\gamma|$ increments linearly from 1 up to a cap of $2\%$ of the total variables, at which point it resets to 1.
This cyclic approach ensures that the system is tested across a spectrum of conditioning densities, ranging from lightly to moderately constrained subspaces.

The content of the evidence $\gamma$ is derived from the seed model $\omega$ but treated differently depending on the query type:
\begin{itemize}
    \item \textbf{Model Counting, Sampling, and Direct Access:} For these tasks, we strictly select $|\gamma|$ literals from $\omega$ without modification. This guarantees that the defined subspace contains at least one solution (the seed itself), preventing trivial zero-count results and ensuring the reasoner must traverse a valid sub-circuit.
    For \emph{uniform sampling} and \emph{direct access}, the query index $i$ denotes the position of the query in the batch $(q_1,\ldots,q_n)$: the $i$-th sampling query requests $i$ samples, while the $i$-th direct access query requests the $i$-th model in the lexicographical order.
    \item \textbf{Satisfiability:} To generate a mix of satisfiable and unsatisfiable queries, we select $|\gamma|$ literals from $\omega$ but randomly flip the polarity of each literal with a probability of $0.5$. This ``perturbed'' evidence tests the solver's ability to distinguish between consistent subspaces and those that have been rendered inconsistent by conflicting constraints.
\end{itemize}

\begin{figure}[t]
  \includegraphics[width=0.472\textwidth]{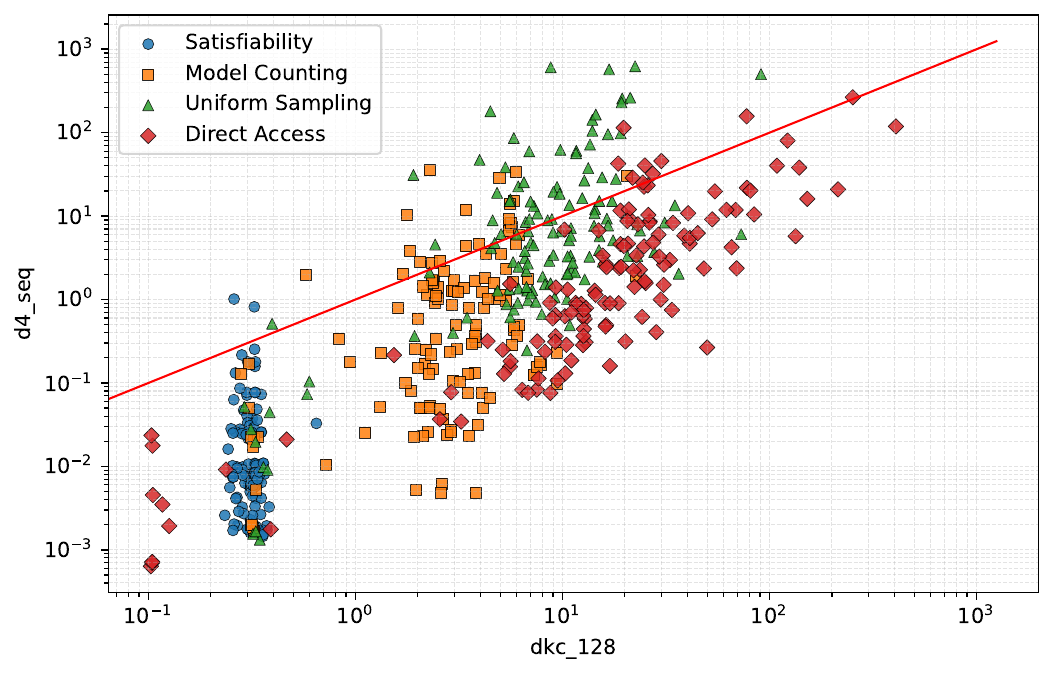}
  \caption{Scatter plot comparing the 4 different queries times of \dkc{} with 128 cores ($x$-axis) versus the sequential baseline \dfour{} ($y$-axis). 
  Points above the diagonal indicate instances where the distributed approach outperforms the sequential one.\label{fig:scatterD4vsDkcQuery}}
\end{figure}

First, we observe that these benchmarks are relatively tractable, as detailed in the supplementary material.
Specifically, the sequential version of \dfour{} successfully compiled 172 instances, whereas \dkc{} solved 181.
In 116 cases, both approaches require less than one second to complete.
For such lightweight instances, the distributed architecture is less advantageous; the inherent communication overhead dominates the runtime and cannot be effectively amortized.
However, on harder instances, \dkc{} outperforms the sequential baseline.
Since these results align with the trends observed in the previous section, we do not discuss them further.
This also reflects the total offline-plus-online cost: \dfour{} is preferable on very small instances, whereas the distributed approach becomes beneficial, and sometimes necessary, when sequential compilation reaches the timeout or memory limit.
Instead, the remainder of this evaluation focuses on the instances solved by both approaches, allowing for a direct comparison of the reasoning engine's efficiency.

Figure~\ref{fig:scatterD4vsDkcQuery} presents a detailed scatter plot comparing the runtime of the sequential baseline \dfour{} against the distributed \dkc{} configuration (128 cores).
In this analysis, we decompose the workload to isolate the performance characteristics of each query; each data point represents the \emph{cumulative} wall-clock time required to answer all queries of a specific type for a single instance.
The results reveal a nuanced landscape where the benefits of parallelism are governed by the computation-to-communication ratio.
In contrast to the compilation phase, where \dkc{} demonstrated clear superiority, the query processing phase exhibits a different trend: a significant portion of data points lie below the diagonal.
This indicates that for these tractable benchmarks, the inherent overhead of the distributed architecture often masks its computational advantages.
However, this result must be contextualized.
Since the reported times aggregate 1,000 queries, the average processing time per query is often sub-second.
In this low-latency regime, network communication becomes the dominant bottleneck.

More specifically, regarding \emph{satisfiability} (blue circles), we observe that the entire batch of queries is typically handled in under 0.1 seconds.
In this ultra-low latency regime, the distributed reasoner cannot compete; the mandatory overhead of broadcasting requests and reducing results dominates the execution time, rendering parallelism ineffective.

For \emph{model counting} (orange squares), the landscape is more nuanced.
While a majority of points remain below the diagonal, we observe a shift for instances where the sequential solving time exceeds 1 second.
In these cases, the local aggregation workload on each worker is substantial enough to mask the communication latency.
Because the time is spent on local computation rather than network synchronization, the distributed approach begins to yield a net benefit.

The most significant performance penalty is observed for \emph{direct access} (red diamonds).
As detailed in Algorithm~\ref{alg:dreasoner_logic}, retrieving a single model requires a sequence of model counting queries (one per variable).
For a formula with $N$ variables, this necessitates $N$ distinct round-trip communications between the Master and the cluster.
Since the individual counting operations are extremely fast, the total runtime is dominated by the accumulation of these synchronization barriers.
Consequently, scaling to 128 cores yields diminishing returns, as the latency of synchronizing a larger cluster exacerbates the cumulative delay.

In contrast, the architecture demonstrates robustness for \emph{uniform sampling} (green triangles), where the performance distribution is significantly more favorable than direct access.
This validates our routing-based strategy: once the initial counts are cached, the Master directs sampling requests to specific workers rather than broadcasting every step.
This unicast mechanism minimizes global synchronization overhead, allowing \dkc{} to effectively leverage the cluster's aggregate memory bandwidth.

In summary, while \dkc{} incurs overhead for lightweight queries on tractable instances, it offers a scalable solution for memory-intensive reasoning tasks where the size of the compiled circuit precludes single-node execution.

\section{Conclusion and Perspectives\label{conclusion}}
In this work, we presented \dreasoner{}, a novel distributed architecture that unifies high-performance knowledge compilation with real-time query processing.
By leveraging the Cube-and-Conquer paradigm, our approach decomposes the search space into disjoint fragments, constructing an implicit global d-DNNF across a cluster of worker nodes.
Unlike traditional distributed model counters that terminate after a single counting operation, \dkc{} transitions into a persistent state, transforming the cluster into a query engine.

Our experimental evaluation clearly demonstrates that \dreasoner{} achieves robust scalability up to 128 cores.
While communication overhead renders the distributed approach less effective for trivial instances, it significantly outperforms the state-of-the-art sequential solver \dfour{} on computationally intensive benchmarks.
Notably, the distributed architecture surpasses the sequential baseline with as few as two active workers.
Furthermore, despite the inherent network latency, we showed that the system efficiently handles complex member queries, such as uniform sampling and direct access, by caching intermediate counts and amortizing the compilation cost across batched requests.

Future work will focus on extending the query language to support weighted model counting, thereby broadening the applicability of \dreasoner{} to probabilistic reasoning.
Additionally, we aim to optimize the latency of iterative queries like direct access, where network round-trips currently dominate the execution time.
We plan to investigate protocol optimizations inspired by \emph{branch prediction} mechanisms.
By employing speculative execution or lookahead strategies, the Master could predict the likely traversal path and speculatively resolve subsequent variable assignments, effectively reducing message volume and masking communication latency.
Finally, we will explore the integration of GPU-accelerated arithmetic circuit representations~\cite{DBLP:conf/iclr/MaeneDM25} to enhance throughput.

\section*{Acknowledgments}

This work has been partly supported by the CERADOC project of the French National Agency for Research (ANR-25-CE23-3078), Jilin province philosophy and social sciences project under Grant 2023ZD15.

\section*{AI Declaration}

The authors declare that no generative AI tools or AI-assisted technologies were used in writing, editing, revising, or preparing this manuscript.

\bibliographystyle{kr}
\bibliography{kr-sample}

\end{document}